\documentclass[twocolumn]{aastex6}

\usepackage{natbib}
\usepackage{graphicx}
\usepackage{booktabs}
\usepackage{longtable}
\usepackage{color}
\usepackage{breakurl}

\shortauthors{Schinzel et al.}
\shorttitle{Radio 3FGL Counterparts}
\slugcomment{final.1.0}

\begin{document}


\title{Radio Follow-up on all Unassociated Gamma-ray Sources from the Third Fermi Large Area Telescope Source Catalog}


\author{Frank K. Schinzel\altaffilmark{1,2}}
\affil{National Radio Astronomy Observatory, P.O. Box O, Socorro, NM 87801, USA}

\author{Leonid Petrov}
\affil{Astrogeo Center, Falls Church, VA 22043, USA}

\author{Gregory B. Taylor\altaffilmark{3}}
\affil{Department of Physics and Astronomy, University of New Mexico, Albuquerque, NM 87131, USA}

\author{Philip G. Edwards}
\affil{CSIRO Astronomy \& Space Science, PO Box 76, Epping 1710 NSW, Australia}



\altaffiltext{1}{fschinze@nrao.edu}
\altaffiltext{2}{An Adjunct Professor at the University of New Mexico.}
\altaffiltext{3}{An Adjunct Astronomer at the National Radio Astronomy Observatory.}

\begin{abstract}
The third \textit{Fermi} Large Area Telescope (LAT) $\gamma$-ray source catalog 
(3FGL) contains over 1000 objects for which there is no known counterpart at 
other wavelengths. The physical origin of the $\gamma$-ray emission of those 
objects is unknown. Such objects are commonly referred to as unassociated and 
mostly do not exhibit significant $\gamma$-ray flux variability. We performed a 
survey of all unassociated $\gamma$-ray sources found in 3FGL using the 
Australia Telescope Compact Array and Very Large Array in the range of 
4.0--10.0\,GHz. We found 2097 radio candidates for association with $\gamma$-ray 
sources. The follow-up with very long baseline interferometry for a subset of 
those candidates yielded 142 new AGN associations with $\gamma$-ray sources, 
provided alternative associations for 7 objects, and improved positions for 
another 144 known associations to the  milliarcsecond level of accuracy. In addition, for 245 
unassociated $\gamma$-ray sources we did not find a single compact radio source 
above 2\,mJy within 3$\sigma$ of their $\gamma$-ray localization. A significant 
fraction of these empty fields, 39\%, are located away from the galactic plane. 
We also found 36 extended radio sources that are candidates for association with a 
corresponding $\gamma$-ray object, 19 of which are most likely supernova remnants 
or HII regions, whereas 17 could be radio galaxies. 
\end{abstract}




\section{Introduction} \label{sec:intro}

The latest generation of $\gamma$-ray observatories, in particular the Large 
Area Telescope (LAT) aboard the \textit{Fermi} satellite have revolutionized
the field of $\gamma$-ray astronomy. The previous major satellite mission,  
the Energetic Gamma-Ray Experiment Telescope (EGRET) on board the Compton-Gamma 
Ray Observatory (CGRO) only detected a couple hundred point sources 
\citep{1999ApJS..123...79H,2008A&A...489..849C}. Since its launch in August 2008, \textit{Fermi}/LAT has 
detected thousands of distinct $\gamma$-ray emitting objects which have been reported 
in multiple catalogs. For all of these catalogs a significant 
fraction, typically $>30\%$, of the detected point sources have no known 
counterpart at any other wavelength and their nature is unknown. This makes
the $\gamma$-ray sky the least understood in all of observational astronomy. 

There are two ways to associate objects from different catalogs: 
1)~to match light curves if a source manifests variability; 2)~to associate
by proximity. Due to errors in the measurement of flux or localization, both methods 
are statistical. Association by proximity is based on two probabilities: the
probability that entries in two catalogs are related to the same object
with the differences in positions due to uncertainties in localization,
and the probability that the second catalog contains a coincident background object,
unrelated to the object in the first catalog. The second probability is crucial
for effective association. If the catalog selected for association contains too
many objects such that the probability to find a background object within the 
localization error ellipse is not small, then the effectiveness of association
with such a catalog will be too low to be practical.

The method for source associations used in the 3FGL is addressed in section 5 
of \citep[3FGL;][]{2015ApJS..218...23A}. A detailed description for the method 
can be found in \citet{2012ApJ...753...83A}. We found in our previous work \citep{2013MNRAS.432.1294P,2015ApJS..217....4S} 
that association based on proximity of \textit{Fermi} sources
to radio sources with emission at 8~GHz at parsec scales is very effective. 
In this paper we will present the results from our method that is complimentary 
to the approach taken by the \textit{Fermi} science team. Following the statistical
criteria described in \citet{2013MNRAS.432.1294P}, the detection of a compact radio 
source brighter than 12\,mJy at 8\,GHz found within the 2-$\sigma$ 
\textit{Fermi} localization error ellipse establishes an association. More than 
50\% of all \textit{Fermi} associations are made
on the basis of this approach. There are two reasons why this association method
based on VLBI detection is so powerful. Firstly, the number of compact radio source
is rather limited: according to the $\log N$--$\log S$ distribution, there are only $\! 100,000$
sources brighter than 10~mJy at parsec scales at 8~GHz \citep{2013MNRAS.432.1294P}. Secondly, variability of both
\textit{Fermi} $\gamma$-ray sources and compact radio sources suggests that emission
is contemporaneous and comes from regions on parsec scales. In contrast, catalogs
from connected interferometers, such as NVSS \citep{1998AJ....115.1693C} and 
SUMSS \citep{1999AJ....117.1578B,2003MNRAS.342.1117M}, typically probe emission at kiloparsec
scales that is related to the interaction of particles in the jet with surrounding media
thousands of years ago.

Our approach is first to observe with connected interferometers fields within
the \textit{Fermi} localization error ellipse at 5--9~GHz and then follow-up detected sources with VLBI. At these
frequencies emission from a compact core usually dominates in contrast to
frequencies below 2~GHz where emission from extended jets or radio lobes generally
dominates. Therefore detection of a source at 5--9~GHz at arcsecond scales
increases the probability of finding detectable emission at milliarcsecond
scales at 8~GHz.  This contrasts to the case when only emission at 0.8--1.4~GHz at 
arcsecond scale is known, which generally does not result in detection of
compact emission. Observations with connected interferometers at 
5--9~GHz serve as a screening tool for potential targets for follow-up 
VLBI observations to find new associations.

In 2012, we started a campaign to image {\it all} regions covered by the \textit{Fermi} localization error
ellipse containing unassociated \textit{Fermi} sources first with connected interferometers at 4--10~GHz,
then with Very Long Baseline Interferometry (VLBI). The results covering the second \textit{Fermi} point source 
catalog \citep[2FGL;][]{2013ApJS..208...17A} were reported in 
\citet{2013MNRAS.432.1294P} and \citet{2015ApJS..217....4S}. The 2FGL presented
1873 sources of which 575 were considered unassociated. Follow-up on those unassociated sources
revealed 865 radio sources at arcsec scales as candidates for association.
We then obtained new associations using VLBI for 76 of the unassociated $\gamma$-ray sources with 
radio-loud active galactic nuclei (AGN). We found that 129 
out of 588 observed $\gamma$-ray sources at arcmin scales did not have a single radio 
continuum source detected above our sensitivity limit within the 
3$\sigma$ $\gamma$-ray localization. These ``empty'' fields were found to 
be particularly concentrated at low Galactic latitudes. 

Since then, a third source catalog was released by the \textit{Fermi}/LAT collaboration 
\citep[3FGL;][]{2015ApJS..218...23A} covering the first 4 years of operations and 
listing 3033 sources among which 1010 were reported to have no plausible 
counterpart at other wavelengths. Most of the identified or associated 
$\gamma$-ray sources are active galaxies. Here we present an update
of our work that includes radio observations for new 3FGL unassociated sources
and presents them in the context of our previous work. 

The 2FGL catalog covers the first two years of the \textit{Fermi}/LAT mission and lists
all point sources found over that time period with about 5$\sigma$ sensitivity. 
At the time of publication it listed 575 unassociated point sources. However, 
since its release the catalog was modified and the version released on May 
18, 2015 contains 651 unassociated $\gamma$-ray sources with a 
median $\gamma$-ray flux of $1.21\times10^{-12}$\,ph\,cm$^{-2}$\,s$^{-1}$ and 
$\gamma$-ray spectral index of 2.31. In 3FGL, covering the first four years
of the \textit{Fermi}/LAT mission, the number of listed unassociated sources as of
May 18, 2015 grew to 1,010 with a median $\gamma$-ray flux of 
$6.63\times10^{-13}$\,ph\,cm$^{-2}$\,s$^{-1}$ and $\gamma$-ray spectral index of 
2.37. Cross-referencing the catalogs, only 300 unassociated $\gamma$-ray sources 
from 2FGL are represented in 3FGL, which leaves 710 new unassociated $\gamma$-ray
sources that have a median $\gamma$-ray flux of $8.96\times10^{-13}$\,ph\,cm$^{-2}$\,s$^{-1}$
and index of 2.31. 

In Section~\ref{sec:obs} new radio observations are described that have been
performed between 2014 and 2015. This is followed by Section~\ref{sec:results}
where we describe the observational results and provide the complete catalog of
radio counterpart candidates for all 3FGL unassociated sources. This is followed 
by a discussion of the subclass of steady $\gamma$-ray emitters in 
Section~\ref{sec:steady}, which are half of all the detected $\gamma$-ray 
point sources. In Section~\ref{sec:discussion} we discuss 
the results in the context of understanding the population 
of unassociated $\gamma$-ray sources. Finally in Section~\ref{sec:steady} we 
provide a summary and conclusions for our reported findings.

\section{Observations} \label{sec:obs}

The 3FGL catalog covers the entire sky, thus we performed follow-up observations
at two radio interferometric arrays: The Australia Telescope Compact Array (ATCA) 
in the southern hemisphere for observing sources with declinations in a range of 
$[-90^\circ, +10^\circ]$ and the Jansky Very Large Array (VLA) in the northern 
hemisphere for observing sources with declinations $[0^\circ, +90^\circ]$. The
overlap in sky area was used to cross-check VLA and ATCA calibration procedures.
Where a source was detected in both ATCA and VLA data, the VLA results were used.

Combining all observing campaigns with VLA and ATCA, we observed 960 fields around 
{\it all} \textit{Fermi} unassociated sources. We excluded 80 sources for which we 
had previously found an association from our analysis of radio sources, which fall within the 
\textit{Fermi} localization error ellipse of 3FGL sources and exhibit parsec 
scale emission.

\subsection{Australia Telescope Compact Array}

We have identified 713 objects marked as unassociated in the 3FGL catalog with 
declinations $<+10^\circ$. Together with our previous observations we were 
able to associate 80 objects with AGN found to have radio emission on parsec scales 
using the method described in \citet{2013MNRAS.432.1294P} and 
\citet{2015ApJS..217....4S}. This is based on the computation of the likelihood
ratio between the probabilities to find an unrelated background source within 
the \textit{Fermi} localization error ellipse and the probability to find a radio 
counterpart within a given
position difference due to random localization errors. The remaining 633 objects
were selected as the primary targets for the observations described here. We 
added 169 secondary targets to the list. These included 122 sources associated 
on the basis of X-ray emission, 14 sources associated with the low frequency catalogs 
NVSS \citep{1998AJ....115.1693C} and SUMSS \citep{2003MNRAS.342.1117M}, 
and 30 unassociated sources that were in the preliminary version of the 3FGL 
catalog but not included in the final release. We also included three sources 
that were marked as unassociated in the preliminary version of the 3FGL but were 
later associated with pulsars. In total, 805 fields with declinations $<+10^\circ$ 
were in the target list.

Observations were made in three campaigns: A3 started on 2014 April 07 and lasted 
for 30 hours, A4 that started on 2014 September 23 and lasted for 66 hours, and 
A5 that started on 2015 April 04 and lasted for 8 hours. Campaign A3 was observed 
in array configuration H168 with baselines ranging from 61--192\,m between the 
inner five antennas and $\sim$4.4\,km between CA06 and the inner antennas, 
and campaigns A4 and A5 were observed in H214 configuration with baselines ranging 
from 92--247\,m between the inner five 
antennas\footnote{\url{http://www.narrabri.atnf.csiro.au/operations/array_configurations/configurations.html}}. 
Observations in all three campaigns were recorded 
simultaneously in two bands centered at 5.5 and 9.0~GHz with a bandwidth of 
2\,GHz in both linear polarizations, with circular polarizations derived
during processing. The primary flux calibrator was PKS\,1934-638. A summary of 
all observations is listed in Table~\ref{tab:observations}.

A list of 405 target sources were observed in the A3 campaign. In the A4 campaign
we observed the remaining 228 primary targets, 172 secondary targets and we
re-observed 303 fields observed in the A3 campaign. There are two reasons for
re-observations: 1) we detected a radio source within the $3\sigma$ $\textit{Fermi}$ 
localization error ellipse, but at a distance of more than $4'$ from the pointing 
direction -- the distance where the primary beam power drops below 20\%
with respect to the pointing direction at 9 GHz (205 fields); and 2) no source 
was detected (98 fields). In the latter case we re-observed the field in 
a 7-element mosaic mode to search for radio counterparts farther from the
\textit{Fermi} position, since typically the \textit{Fermi} localization error ellipse
is wider than the ATCA field of view. During the A5 campaign we re-observed 13 primary
targets that were missed in A4, 6 secondary targets that 
were recorded only in 1 scan in A4 and re-observed 120 fields where a source
was found at distance more than $4'$ from the pointing direction.

   Among primary targets 45\% of the fields were observed in 3 scans of 24\,s each, 
39\% were observed in 4 scans, and 16\% were observed in 5 scans. Among
secondary targets, 37\% of the sources were observed in 2 scans; other sources 
were observed in 3 or more scans.

\begin{figure}
   \centering
   \includegraphics[width=0.48\textwidth]{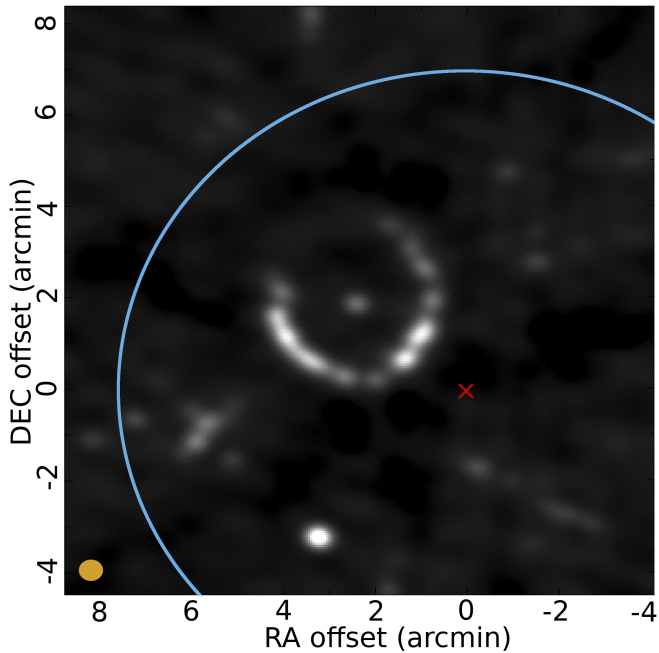}
   \caption{Source J1811-1925 located in the heart of the young 
            supernova remnant G11.2--0.3. The position of the sources in the center 
            of the nebula is within $7''$ of the X-ray pulsar PSR J1811-1925 
            \citep{2004AdSpR..33..592G}. The \textit{Fermi} position is shown with 
            a cross and the ellipse marks the edge of the 95\% confidence of the
            $\gamma$-ray localization. According to \citet{1977hisu.book.....C}, 
            the remnant is plausibly associated with the historical "guest star" 
            witnessed by 
            Chinese astronomers in the year 386 A.D. The \textit{Fermi} source 
            J1811.3-1927c lies at a distance of 2.9$'$ from the radio source. The 
            bright source south of J1811-1925 is J1811-1930 for which no parsec
            scale emission was detected above 6\,mJy. The filled circle in the 
            lower left corner indicates the size of the synthesized beam. The 
            image shows 5.5 GHz observed on 2014 Sep. 24 with ATCA. The image 
            rms at the center of the field of view is 2.2~mJy and the peak 
            flux density in this field 252.7\,mJy.}
   \label{f:aofus_ima1}
\end{figure}

\begin{figure}
   \centering
   \includegraphics[width=0.48\textwidth]{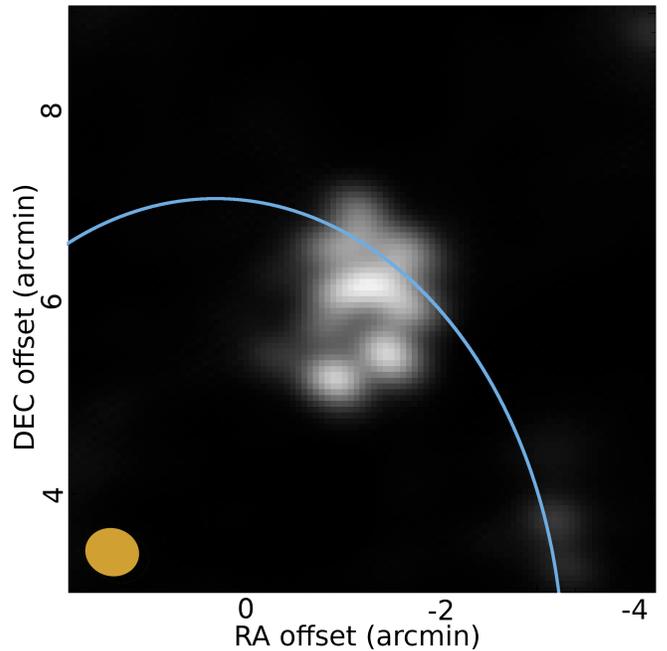}
   \caption{Source J1814-1755 in the ultra-compact HII region 
            W33, also known as G012.8050-00.2007 in the CORNISH 
            catalog \citep{2013MNRAS.435..400U}. The nearby $\gamma$-ray
            source is 3FGL\,J1814.0-1757c. The edge of its 95\% confidence
            localization ellipse is indicated. The filled circle in the 
            lower left corner indicates the size of the synthesized beam.
            The image shows 5.5 GHz observed on 2014 Sep. 24 with ATCA. The 
            image rms in the center of the field of view is 0.7~mJy and the
            peak flux density in this field is larger than 40\,mJy.
           } 
   \label{f:aofus_ima2}
\end{figure}

   The data analysis procedure we used for the A3, A4, and A5 campaigns
is described in detail in \citet{2013MNRAS.432.1294P}. We obtained radio images 
of size  $1024 \times 1024$  with a pixel size of 2.4~arcsec and a 
synthesized beam with a typical FWHM size $20'' \times 30''$. The majority of 
sources found look point-like. Two examples of images with extended structure are 
shown in Figures~\ref{f:aofus_ima1} and \ref{f:aofus_ima2} \footnote{Compare the 
image with \url{http://cornish.leeds.ac.uk/cgi-bin/public/summary\_src.py?name=G012.8050-00.2007},
\url{http://atlasgal.mpifr-bonn.mpg.de/cgi-bin/ATLASGAL\_SEARCH\_RESULTS.cgi?text\_field\_1=G012.8057-0.1994\&catalog\_field=GaussClump}}, 
though these are the rare exceptions. We determined the flux density of the 
detected sources, spectral index within a band, and positions. Typical 
position uncertainty is around $1''$ and typical uncertainty in flux 
density is 0.1--0.4~mJy. The detection limit is 1.0--1.5~mJy for sources in the 
center of the field-of-view. For sources detected at both 5.5 and 9.0~GHz 
bands we determined the spectral index across the bands.

   Since target sources are observed in 2--3 scans of 18\,s length with 
a 6-element interferometer, the estimates of flux densities of extended objects 
may have a strong bias and underestimate the peak flux density by up to one order 
of magnitude. Scrutinizing the data, we found 1--2\% of the reported objects
to have questionable flux density estimates. The poor sampling of the Fourier 
plane due to snapshot observations 
makes it difficult to filter outliers reliably. When in doubt, we tended to keep 
questionable flux density estimates,
considering it better to keep problematic flux density estimates than to
eliminate a potential candidate for association. Follow-up VLBI observations that 
are sensitive to the compact components of the objects will filter out galactic objects 
that do not have detectable emission on milliarcsecond scales. 

   We found a radio object within at least one band and within the area of the 3$\sigma$ \textit{Fermi}
localization error ellipse of 497 3FGL unassociated sources out 
of 633 (79\%). For the remaining 21\% of the sources no radio counterpart 
brighter than 2~mJy was found.

\subsection{Very Large Array}

A list of 322 unassociated 3FGL fields with declinations above 0$^\circ$ were 
selected for observations with NRAO's Jansky Very Large Array (VLA) in this 
campaign (V2). Additionally, we observed the location of 2FGL\,J0423.4+5612, for 
which no data was recorded in our previous VLA survey. We reanalyzed our previous 
campaign V1 \citep{2015ApJS..217....4S} and found that 51 out of 169 (30\%) of observed fields lie within $4'$ of 3FGL 
sources. Among these 51 sources, 18 are marked as associated in 3FGL, and 5 others 
were observed in V2. Therefore, combining V1 and V2, we get 491 fields. Of these, 
327 are centered at 3FGL sources marked as unassociated, 18 are marked as 
associated, and 146 are pointed at directions farther than $4'$ from a 3FGL source.

The observations were conducted using
the C-Band receiver covering a frequency range of 4--8\,GHz, recording a total
bandwidth of 4\,GHz in both left and right handed circular polarizations. The
VLA array was in B configuration, thus an integration time of 2\,s was chosen.
Real-time correlation was performed using VLA's WIDAR (Wideband Interferometric
Digital ARchitecture) correlator. The instantaneous bandwidth was split into 
two parts, with one half centered at 5.0 GHz (4.0--6.0\,GHz) and the other 
centered at 7.0 GHz (6.0--8.0\,GHz). This provides a simultaneous observation of 
two separate frequency bands. The observing time of 10 hours was split into five 
segments to be able to observe fields at all LST ranges and to increase 
the likelihood of scheduling (see Table~\ref{tab:observations}). The first four 
segments were observed between 2015 March 16 and 21 under time approved 
through the NASA Fermi Guest Investigator program, an additional hour to 
complete the program was approved as director's discretionary time
observed on 2015 April 16.

At the beginning of each observing segment either 3C\,48 or 3C\,286 was 
observed to provide a bright flux
density/bandpass calibrator. In most cases each target source was observed only once with
a total integration time of $\sim$30\,s each. Nearby phase calibrators
were added with typical integration times of 15\,s each in order to 
solve for changes in the complex gains during the target
observations.  For all segments the VLA was in B array configuration,
providing baseline lengths from 0.21 to 11\,km, which results in
1\,arcsecond resolutions and a field-of-view of up to about 7$'$. The
largest angular scale of extended radio structures is about 29$''$.

The data calibration and analysis procedure is similar to what is described in 
\citet{2015ApJS..217....4S}. Major differences were that the newer Common Astronomy 
Software Applications (CASA) release 4.5.2 and the 31DEC2015 release of the
the Astronomical Image Processing System \citep[AIPS;][]{1990apaa.conf..125G} 
were used.\footnote{for more information on CASA see
\url{http://casa.nrao.edu/} and for AIPS see \url{http://www.aips.nrao.edu/index.shtml}}
The data calibration was performed entirely within CASA. 

The imaging was performed in CASA using the \texttt{clean} task. This task uses 
a Clark based clean algorithm \citep{1980A&A....89..377C}, applies the $w$-projection 
needed for widefield imaging using 100 projection planes, and visibility weights
are determined using the Briggs weighting scheme \citep{PhDT...1995B},
minimizing sidelobes and noise levels.  We used ``$\mathrm{robust}=0$'', which
corresponds to a weighting scheme in between uniform and natural. In addition,
we used multifrequency synthesis with two terms in order to compensate for 
spectral changes over the 2\,GHz instantaneous bandwidth. This provides both 
a Stokes I map combining the two polarizations together and a spectral index
map. The deconvolution was run with 5000 iterations, a default loop gain of 0.1, 
and a flux density threshold at which to stop cleaning of 0.06\,mJy corresponding 
to the approximate thermal noise limit. The primary beam attenuation was corrected
using new measurements described in \citet{PerleyEVLAMemo195}. 

All images were then searched using an automatic procedure reading images into 
AIPS and using the task \texttt{SAD} to identify point sources. Custom python 
scripts were applied to analyze and inspect the resulting images and point sources
found. Sometimes the algorithm applied in \texttt{SAD} identifies image artifacts 
near strong point-sources, which were manually flagged in order to retain 
a clean list of point sources that includes position, flux density, and spectral
index for each of the two 2\,GHz sub-bands which are used in subsequent analysis.

\begin{table*}[htbp!]
    \centering
    \caption{List of observations with connected interferometers.\label{tab:observations}}
    \small
  \begin{tabular}{llcccccll}
    \hline
     T & C & Code & Start & Dur.     & Tune & B/w    & \#  & Identifier\\
       &   &      & (UTC) & (h)      & (GHz)& (GHz)  &   \\
    \hline
   ATCA    & H214 & C2624   & 2012 Sep 19 10:00 & 29 & 5.5/9.0 & 2.0 & 411 & A1\\
   VLA     & A    & S5272   & 2012 Oct 26 11:19 & 2  & 5.0/7.3 & 1.0 & 41  & V1A\\
   VLA     & A    & S5272   & 2012 Nov 03 21:01 & 7  & 5.0/7.3 & 1.0 & 175 & V1B\\
   ATCA    & H214 & C2624   & 2013 Sep 25 21:30 & 45 & 5.5/9.0 & 2.0 & 997 & A2\\
   ATCA    & H168 & C2624   & 2014 Apr 07 16:00 & 30 & 5.5/9.0 & 2.0 & 405 & A3\\
   ATCA    & H214 & C2624   & 2014 Sep 23 04:00 & 66 & 5.5/9.0 & 2.0 & 703 & A4\\
   VLA     & B    & S7104   & 2015 Mar 16 21:33 & 1 & 5.0/7.0 & 2.0  & 30  & V2A\\
   VLA     & B    & S7104   & 2015 Mar 16 22:33 & 3 & 5.0/7.0 & 2.0  & 107 & V2B\\
   VLA     & B    & S7104   & 2015 Mar 17 09:46 & 1 & 5.0/7.0 & 2.0  & 21  & V2C\\
   VLA     & B    & S7104   & 2015 Mar 21 12:37 & 4 & 5.0/7.0 & 2.0  & 156 & V2D\\
   ATCA    & H214 & C2624   & 2015 Apr 04 04:00 & 8 & 5.5/9.0 & 2.0  & 139 & A5\\
   VLA     & B    & 15A-466 & 2015 Apr 16 20:00 & 1 & 5.0/7.0 & 2.0  & 19  & V2E\\
   \hline
  \end{tabular}
  \begin{flushleft}
    Column description:\\
    T -- telescope\\
    C -- array configuration\\
    Code -- observation proposal code\\
    Start -- start time in UTC\\
    Dur. -- duration of the observation in hours\\
    Tune -- center frequency of the tunings\\
    B/w -- bandwidth of each tuning\\
    \# -- number of targets observed\\
    Identifier -- custom observation identifier\\
  \end{flushleft}
\end{table*}

\subsection{VLBI follow-up observations}

  We should stress that estimates of flux densities of emission at kiloparsec
scales determined from analysis of ATCA and VLA observations are not sufficient
to provide high-confidence associations for \textit{Fermi} sources. Firstly, emission at arcsec
resolutions can have a significant contribution from extended regions of an AGN that often
dominate over emission from the AGN's parsec scale core. Variability time scales of $\gamma$-ray fluxes
strongly suggest $\gamma$-ray emission to be generated at parsec scales.
Secondly, the number of sources with flux density at arcsecond resolution
above some limit is significantly greater than the number of sources with flux 
density above the same limit at milliarcsecond resolutions. Given the large number
of weak radio sources located within a typical \textit{Fermi} localization error
ellipse at arcsecond resolution, this prevents association of sources weaker
than 30--50~mJy. Thus, we suggest that ATCA and VLA observations provide only 
initial candidates for associations.

  Here we report results of our follow-up observations of candidate sources with VLBI. 
We included 144 target sources 
detected in ATCA and VLA observations in the ongoing VLBA Calibrator Survey 
Densification campaign (VCS8/9) at 4.4 and 7.6~GHz \citep{r:vcs9}\footnote{\url{http://astrogeo.org/vcs9}}. 
The goal of this program is to increase the density of calibrator sources. New AGN 
candidates for 
$\gamma$-ray association turn out to be good targets for this
program. We included a number of target sources in the ongoing LBA Calibrator 
Survey (LCS-2) at 8.4~GHz that has similar goals as the VLBA program but is 
focused on sources in the southern hemisphere. The data analysis procedures of 
the observations is similar to previous VCS and LCS campaigns and is
described in \citet{2008AJ....136..580P} and \citet{2011MNRAS.414.2528P}. 
The detection limit of these programs was in a range of 10--12~mJy for
sources at elevations above $20^\circ$ at the telescopes.

  We also ran a dedicated VLBA program at 7.6\,GHz in 2015 and in 2016 for 
observing 561 target sources detected by the ATCA and VLA (codes BS241 and S7104)
with flux density at 9\,GHz exceeding 10\,mJy or at 5.5/7\,GHz exceeding 20\,mJy. 
Target sources were observed in two scans of four minutes each using
a bandwidth of 480~MHz, and dual polarization. We ran two LBA experiments 
in 2014 and 2016 at 8.4~GHz for observing 140 target sources at declinations 
below -$40^\circ$. These target sources were observed in two scans of 210~s each
within the spanned bandwidth of 320~MHz. The detection limit for dedicated VLBI 
observations is 6~mJy when a source is at elevations above $20^\circ$ at 
the telescopes. All VLBI observations were correlated with accumulation 
periods of 0.1~s and a spectral resolution of 62.5--125~kHz in order to have a wide 
field-of-view with sensitivity reduced by no more than 20\% within $1'$ of 
the pointing direction.

  Additional VLBI observations to cover the remaining $\gamma$-ray counterparts
brighter than 10\,mJy in any band are ongoing and will be presented 
in a future publication.

\pagebreak

\section{Results} \label{sec:results}

\subsection{Catalog of Radio Candidates}

Similar to \citet{2015ApJS..217....4S}, we combined the results from the ATCA and 
VLA observing campaigns into a single uniform dataset. We combined results of 
three ATCA campaigns, A3, A4, A5, the VLA campaign V2 that we described in the 
previous sections and two ATCA and one VLA campaign. This continues our 
program of observing unassociated
sources in 1FGL and 2FGL catalogs described in our previous publications 
\citep{2015ApJS..217....4S,2013MNRAS.432.1294P}, A1, A2, and V1. We retained
2097 sources, 50\% of the total number of detected sources, that are 
within the $3\sigma$ \textit{Fermi} localization error ellipse and have flux densities above 1~mJy
in at least one band. The catalog presented here for the fields observed in 
the previous A1, A2, and V1 campaigns are not identical to those previously published
and is only valid for 3FGL localizations.
Firstly, we re-analyzed the observations. Secondly, since positions of 
$\gamma$-ray sources in 3FGL and their uncertainties are slightly different than 
positions in the 1FGL and 2FGL catalogs, some radio sources absent in the published 
versions are now closer to the 3FGL positions than to 1FGL or 2FGL and are now 
included, while some radio sources with positions beyond the 3FGL 3$\sigma$ \textit{Fermi} 
localization error ellipse are now excluded. 

If a source was detected at distances from the pointing direction beyond where 
the total power drops to 20\% of the pointing center, which is $6.2'$ at 
5~GHz and $4.2'$ at 7~GHz, we
report only a lower limit on the flux density. The detection limit at the center of the 
field-of-view is 1~mJy and is 5~mJy at the edge. A number of sources are detected 
only at one band. This may be due to the source spectrum or because a source
was too distant from the pointing direction. We report spectral index 
estimates only if a source was detected at both bands within $4.2'$ of the pointing 
direction.

All radio sources found within the 3-$\sigma$ \textit{Fermi} 
localization error ellipse were cross-matched against the reprocessed TIFR GMRT 
150 MHz Sky Survey (TGSS) alternative data release 1 \citep[ADR1:][]{2016arXiv160304368I}, 
NRAO VLA Sky Survey \citep[NVSS:][]{1998AJ....115.1693C}, Sydney University Molonglo Sky Survey 
\citep[SUMSS, version 2.1 of 2012 February 16:][]{1999AJ....117.1578B,2003MNRAS.342.1117M}, 
the Molonglo Galactic Plane Survey 2nd Epoch \citep[MGPS-2:][]{2007MNRAS.382..382M}, 
the Gaia Data release 1 \citep[Gaia:][]{2016arXiv160904172G}, and the Wide-field Infrared 
Survey Explorer (WISE) catalog\footnote{\url{http://wise2.ipac.caltech.edu/docs/release/allwise/}}
\citep[ALLWISE, November 13, 2013:][]{2010AJ....140.1868W,2011ApJ...731...53M}, 
which combines the data from the WISE cryogenic and post-cryogenic survey phases 
providing the most comprehensive view of the full mid-infrared sky currently 
available. We obtained 837 matches with the ALLWISE catalog, 1291 matches with 
the Gaia DR1 catalogue, 1051 matches with NVSS, 461 matches with SUMSS and MGPS-2, 
and 507 matches with TGSS ADR1. We evaluated the probability of a false detection
for Gaia counterparts in the following way. We counted the number of sources in a uniform 
$0.25^\circ \times 0.25^\circ$ grid and converted it to the density of Gaia sources
per steradian. We then defined the probability of false detection as the product of
the area in the search radius and local Gaia sources density \citep{r:rfc_gaia}. 
The search radius for sources associated with VLBI is $0.2''$, and $3.0''$ 
otherwise. A search radius of $0.2''$ for VLBI associated sources was selected 
to accommodate possible position errors in Gaia.

Table~\ref{t:cat} presents the catalog of detected sources within 3-$\sigma$ of the
\textit{Fermi} localization error ellipse. The table of the remaining 1842 sources 
detected outside of the 3-$\sigma$ \textit{Fermi} localization error ellipse can 
be found in the online attachment.

\begin{table*}
  \caption{The first 8 rows of 2097 objects found within 3$\sigma$ \textit{Fermi} 
           localization error ellipse. The column descriptions are explained in the tables notes. 
           The full table is available in the electronic attachment.}
  \tiny
  \begin{verbatim}
    (1)             (2)          (3)        (4)   (5) (6) (7) (8)  (9) (10) (11)  (12) (13) (14)     (15)           (16) (17)      (18)         (19)    (20)        (21)             (22)  
FRC J0000+6309  00 00 19.26 +63 09 51.96   0.30  0.30    14.8  0.1                           V2 3FGL J0001.0+6314   6.68 1.73                                                              
FRC J0000+6315  00 00 17.36 +63 15 35.93   0.30  0.30     1.3  0.1                           V2 3FGL J0001.0+6314   4.98 1.35                                                              
FRC J0000-3738  00 00 08.39 -37 38 19.95   0.90  0.85    17.4  0.1    17.6  0.2   0.02 0.03  A3 3FGL J0000.2-3738   1.36 0.86  RFC J0000-3738   22.0   206.5  SUMSS J000008-373819    18. 
FRC J0001+3524  00 01 38.83 +35 24 31.24   0.30  0.30              >   3.7  0.1              V2 3FGL J0001.6+3535  10.92 2.24                                                              
FRC J0001-4155  00 01 32.72 -41 55 25.41   0.80  0.80     9.5  0.1    12.0  0.3   0.47 0.07  A4 3FGL J0002.2-4152   8.24 2.52  RFC J0001-4155    9.0     1.3  SUMSS J000133-415524    13.  
FRC J0002+6219  00 02 53.52 +62 19 17.04   0.30  0.30     3.5  0.2     1.8  0.2  -1.76 0.44  V1 3FGL J0002.6+6218   1.84 1.45                                                              
FRC J0002-6716  00 02 28.13 -67 16 11.57   0.80  0.80     4.9  0.1     4.2  0.1  -0.31 0.09  A4 3FGL J0002.0-6722   6.38 2.96                                 SUMSS J000228-671612    32. 
FRC J0002-6726  00 02 15.19 -67 26 53.41   0.80  0.80    11.2  0.1    14.8  0.1   0.57 0.03  A4 3FGL J0002.0-6722   4.78 2.13  RFC J0002-6726                 SUMSS J000215-672652    19. 

     (23)              (24)       (25)                 (26)       (27)                 (28)       (29)                  (30)    (31)            (32)         (33)
NVSS 000019+630951     18.9                                  WISE J000019.04+630952.9 15.72
NVSS 000017+631538      3.0                                  WISE J000017.42+631535.5 14.29  GAIA  430090779421378688 19.98  00 00 17.28832 +63 15 38.4358  0.00076
NVSS 000008-373819     15.4                                  WISE J000008.41-373820.6 14.57
NVSS 000138+352432     10.9                                  WISE J000138.71+352430.1 15.93
                                                             WISE J000132.74-415525.2 13.99  GAIA 4995979729065629312 18.40  00 01 32.75326 -41 55 25.3234  0.00018
NVSS 000253+621917     12.7  TGSS J000253.2+621917    140.4  WISE J000253.17+621917.6 14.19  GAIA  429932308004616192 17.23  00 02 53.22260 +62 19 18.0223  0.00003
                                                             WISE J000228.45-671608.7 14.60
                                                             WISE J000215.19-672653.4 14.51  GAIA 4707413417751868800 18.21  00 02 15.18729 -67 26 53.4893  0.00005

  \end{verbatim}
    Column descriptions: (1) -- shows the IAU name; (2),(3) show right ascension 
    and declination; (4),(5) show uncertainties in right ascension without 
    $\cos \delta$ factor; (6),(7),(8) show flux density flag, flux density 
    estimate at the low frequency band and its uncertainty in mJy; (9),(10),(11) 
    show flux density flag, flux density estimate and its uncertainty at the high 
    frequency band 7/9\,GHz; (12),(13) show spectral index defined as $S \sim \nu^{+\alpha}$ 
    between low and high frequency bands; (14) shows the campaign code; (15) 
    shows 3FGL source name; (16) shows distance between the radio sources and 
    the $\gamma$-ray source position in arcmin; (17) shows the same distance 
    divided by 1-$\sigma$ \textit{Fermi} position uncertainty; (18),(19),(20) 
    show the name of a counterpart in the VLBI catalog RFC\footnote{Available at http://astrogeo.org/rfc}
    (Petrov \& Kovalev, 2017, in preparation) within $20''$, the correlated
    flux density at 8~GHz in mJy, and the likelihood ratio (as described in 
    Section~\ref{sec:new_assoc} of radio-$\gamma$-ray association); (21),(22) 
    show the name of a counterpart from SUMSS or MGPS-2 catalogs found within 
    $40''$ and its flux density in mJy; (23),(24) show the name of a counterpart 
    from NVSS catalog found within $20''$ and its flux density in mJy; (25),(26) 
    show the name of a counterpart from TGSS catalog and flux density at 150~MHz; 
    (27),(28) show the name of a counterpart from ALLWISE catalog found within 
    $3''$ and its magnitude at wavelength 3.4 microns; (29),(30) show an ID of 
    a counterpart from optical Gaia DR1 catalog found within $0.2''$ for sources 
    associated with VLBI and $3.0''$ otherwise and its magnitude at filters G; 
    (31),(32) shows Gaia position; (33) shows the probability of false association
    with Gaia. If the flux density flag in (4) or (7) is '$<$', that means the 
    source was detected with a sidelobe and the reported flux density is the 
    lower limit.
  \label{t:cat}  
\end{table*}

\subsection{New Associations\label{sec:new_assoc}}

In order to establish the association of a parsec scale source with its $\gamma$-ray
counterpart we run a likelihood ratio test using Poisson statistics. The likelihood 
ratio of association $\Lambda$ is defined as the probability that the $\gamma$-ray and 
radio source found at separation $d$ is physically the same object, and their position 
difference is due to statistical errors only, to the probability that the radio 
source is a background, unrelated object. Simple geometric consideration results in
\begin{equation}
  \Lambda = \frac{e^{-n^2/2}}{N\left(1\right) {S^{{}^p}} d^2/4},
\end{equation}\label{e:lik}
with $d$ being the angular separation in radians between the radio and the $\gamma$-ray position, $S$ is the
radio flux density in Jansky, $n$ is the normalized distance between the radio and $\gamma$-ray
source\footnote{The distance between the radio and $\gamma$-ray localization divided by 
the 1$\sigma$ Fermi localization error.}, $N\left(1\right)=374$, and $p=-1.2088$. The numerical
values were determined from $\log N$--$\log S$ relation of correlated flux
densities. A detailed description for determining the association probabilities 
and the numerical value of the $\log N$--$\log S$ diagram are given in 
\citet{2013MNRAS.432.1294P} and \citet{2015ApJS..217....4S}.

  In total, we performed VLBA or LBA observations in the fields that contain 
798 sources detected in our VLA and ATCA programs within $1'$ of the pointing 
direction. We employed such a wide field-of-view because a compact source with 
emission at parsec scales is not always coincident with the peak brightness 
of the radio structure on kiloparsec scales. \citet{r:obrs2} demonstrated a number 
of such examples.

  We have detected 451 compact sources from these VLBI observations.  
Among 2097 sources detected in our ATCA and VLA programs, there are 744 objects 
with flux densities greater 10~mJy in either 5.5 or 9~GHz. Of these, 630 or 85\%
have been followed up with VLBI. The remaining 114 objects, primarily in the southern
hemisphere, will be observed in the near future.

    We have computed the likelihood ratio among 451 VLBI sources detected within 
$1'$ of objects observed in VLA and ATCA programs. We consider the sources with
$\Lambda > 8$ and a normalized arc length of $n< 3$ as associations. The probability
that the normalized arc-length exceeds 3 due to random errors of the $\gamma$-ray 
source position is 1.1\%. That means setting this criteria we will have false 
negative associations for 1\% of the sources. We selected the $\Lambda > 8$ 
criteria to have approximately the same number of false positive associations. 
In total, we have 19 double VLBI associations for 1530 $\gamma$-ray sources. 
Excluding two gravitational lenses 3FGL\,J0221.1+3556 and 3FGL\,J1833.6-2103, we 
get a false positive probability of 1.1\%. We performed an additional test: we 
rotated the VLBI catalog at random angles in a range of [0.2, 100] degrees, ran 
256 test associations, and found on average of 18.9 associations. Thus, we conclude 
that the false association probability of this criteria, both positive and negative, is 
around 1\%.

  We found that 286 sources out of 451 (63\%) have , and therefore 
are listed as associations based on their parsec-scale emission. Of these, 
144 have associations reported in the 3FGL catalog. Of those, 3 \textit{Fermi} 
sources have two associations. Association based on parsec-scale emission 
confirms them and allows us to improve their position accuracy to milliarcsecond 
scales. We have further established new associations with AGNs for 142 sources. 
Of these, 2 \textit{Fermi} sources have two associations.

\begin{figure}
   \includegraphics[width=0.48\textwidth]{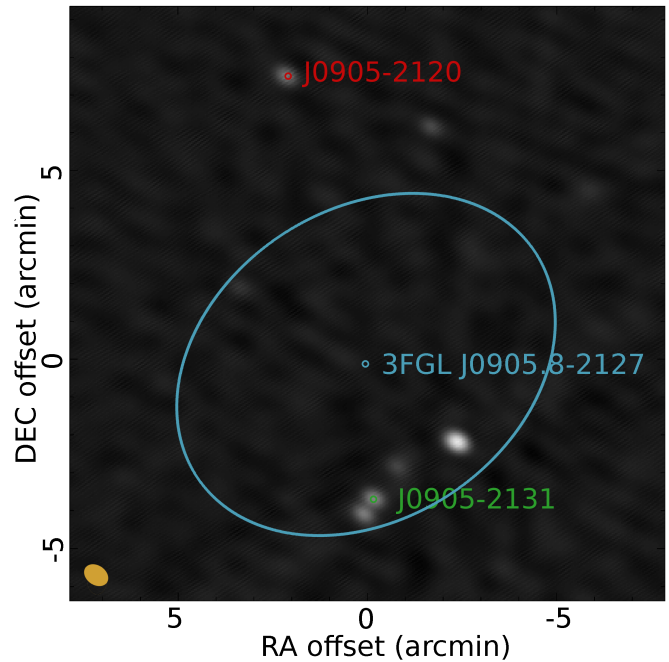}
   \caption{The field observed with ATCA at 5.5 GHz centered on 3FGL J0905.8-2127. 
            Among six point-like sources seen in the field, two have been
            detected with the VLBA at 7.6 and 4.4 GHz. Source {\sf J0905-2131}
            has a likelihood ratio of 22.9, which is above the threshold for
            establishing association based on parsec-scale emission. The filled circle in the 
            lower left corner indicates the size of the synthesized beam. 
            The image shows 5.5 GHz observed on 2014 Apr. 08 with ATCA. The image rms 
            in the center of the field of view is 0.2~mJy and the peak flux density is
            21.9\,mJy.
            }
   \label{f:J0905.8-2127}
\end{figure}

\begin{figure}
   \includegraphics[width=0.48\textwidth]{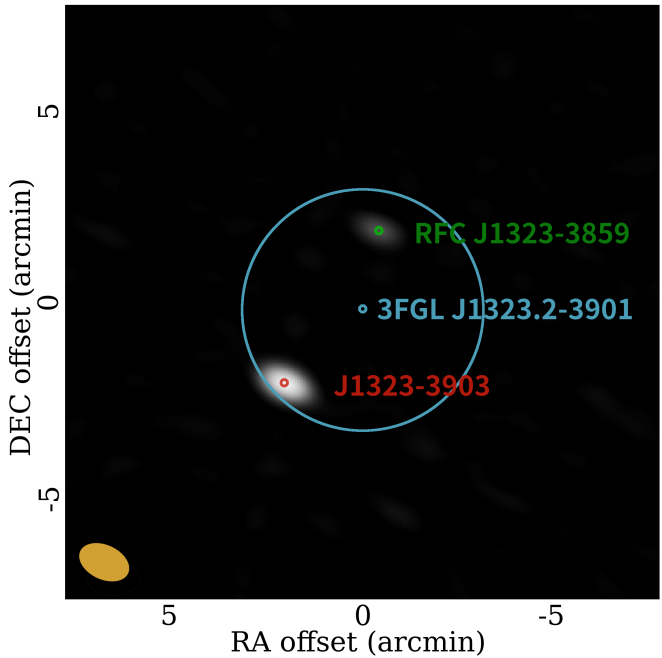}
   \caption{The field observed with ATCA at 5.5 GHz centered on 3FGL\,J1323.2-3901. 
            ATCA observations identified two sources, {\sf J1323-3859} and 
            {\sf J1323-3903} as candidates for association. VLBI follow-up
            observations at 8\,GHz revealed compact emission only from the weakest source 
            {\sf J1323-3859} and thus, discriminated against the other candidate. The filled circle in the 
            lower left corner indicates the size of the synthesized beam.
            The image shows 5.5 GHz observed on 2014 Apr. 24 with ATCA. The image rms 
            in the center of the field of view is 0.1~mJy and the peak flux density is 
            29.0\,mJy.}
   \label{f:3FGL_J1323.2-3901}
\end{figure}

   In the majority of fields observed with ATCA and VLA we found more
than one point source that is a candidate for association. VLBI
observations allow us to find which of these point sources have
parsec-scale detectable emission and therefore identify the likely
counterpart. Figure~\ref{f:J0905.8-2127} provides an 
illustration of this. There are 6 point sources in the field. Of them, two
show parsec scale emission at $>$20~mJy level. The first, {\sf J0905-2120} has 
a $\Lambda$ of 2.5, which is below the threshold
of 8 for a reliable association, and the second {\sf J0905-2131}
which is a double at $28''$ has a $\Lambda$ of 22.9 and thus, we
considered it to be associated with the 3FGL object {\sf J0905.8-2127}.
Figure~\ref{f:3FGL_J1323.2-3901} illustrates another situation. There are two
sources detected with ATCA $4.7'$ apart within the 3FGL\,J1323.2-3901 localization error ellipse: 
{\sf J1323-3859} and {\sf J1323-3903}. The first source has a flux density of 6.3~mJy 
at 5.5~GHz and a spectral index of +0.5, while the second one has flux density of 29.0~mJy 
and a spectral index of $-$0.8. One may expect that the strongest source is a probable 
association. However, VLBI observations revealed compact emission from the 
weaker source {\sf J1323-3859} and no compact emission from 
{\sf J1323-3903}\footnote{It is interesting that the VLBI flux density of {\sf J1323-3903} at 8~GHz, 
23~mJy at epoch 2015.02.24 and 31 mJy at epoch 2015.11.10, is significantly 
higher than its ATCA flux density 8~mJy at 9~GHz at epoch 2014.04.07. The flux 
density of {\sf J1323-3903} found in the same field extrapolated to 1.4 GHz is $89 \pm 6$ which is 
rather close to the value from NVSS of 70~mJy. The flux density of {\sf J1323-3859} 
extrapolated to 1.4~GHz would be $3.1 \pm 0.4$~mJy and it was not detected in the NVSS 
with a limit of 2.4~mJy. These comparisons with NVSS demonstrate 
that there is no gross oversight in the ATCA data calibration. Therefore, we 
attribute the discrepancy between ATCA and VLBI flux density estimates to
source variability.}

  Finally, we present Table~\ref{t:ass} which provides the list of our proposed 
291 associations based on parsec-scale emission at 8~GHz detected with follow-up 
VLBI of point sources found in our ATCA and VLA observations. We note that
association of {\sf J1413-6520} in this list is problematic. The ATCA
observations provide the position of the Circinus galaxy, while with VLBI we 
detected only emission from the supernova remnant 1996cr, $20.86''$ away. 

\begin{table*}
    \caption{The first 8 rows of the list of 291 associations of 3FGL 
             objects that have a likelihood ratio of their association
             of more than 8 based on their parsec-scale radio emission 
             detected with follow-up VLBI observations. The full table 
             is available in the electronic attachment.}
    \par\medskip\par
    \centering
    \scriptsize
    \begin{tabular}{llrrrrrrrrrr}
      \hline
      
      3FGL name & VLBI Name & Right Ascension & Declination & $ \sigma_\alpha $ & $ \sigma_\delta $  & Corr & D & N$\sigma$ & S & 
         $\Lambda$  & 3FGL ass\\
         & & h \hspace{0.1ex} min \hspace{0.5ex} s \hspace{6.0ex}  & 
               $^\circ$ \hspace{0.5ex} $'$ \hspace{0.5ex} $''$ \hspace{5.0ex} & mas & mas & & $'$ & & mJy  &          \\
      \hline
3FGL J0000.2$-$3738 & RFC J0000$-$3738 & 00 00 08.414182 & -37 38 20.67354 &   0.54 &  1.10 &  0.178 & 1.36 & 0.86 &  22.0 &  207.2 &            \\
3FGL J0003.2$-$5246 & RFC J0003$-$5247 & 00 03 19.600260 & -52 47 27.28128 &  19.44 & 10.13 & -0.272 & 1.04 & 0.70 &  15.0 &  194.3 &  RBS 0006  \\
3FGL J0006.2$+$0135 & RFC J0006$+$0136 & 00 06 26.924724 & +01 36 10.38555 &   1.07 &  2.16 & -0.413 & 2.64 & 1.48 &  14.0 &   33.0 &            \\
3FGL J0007.4$+$1742 & RFC J0007$+$1745 & 00 07 18.873920 & +17 45 34.56806 &   0.16 &  0.26 & -0.201 & 3.72 & 0.84 &  22.0 &   28.2 &            \\
3FGL J0007.9$+$4006 & RFC J0007$+$4008 & 00 07 41.666525 & +40 08 29.93735 &   0.42 &  0.66 &  0.258 & 3.43 & 1.19 &  14.0 &   20.1 &            \\
3FGL J0008.3$+$1456 & RFC J0008$+$1456 & 00 08 25.399845 & +14 56 35.79096 &   0.18 &  0.32 & -0.049 & 1.19 & 0.40 &  21.0 &  104.5 &            \\
3FGL J0008.6$-$2340 & RFC J0008$-$233A & 00 08 35.399660 & -23 39 28.00793 &   0.32 &  0.80 &  0.133 & 1.70 & 1.09 &  29.0 &  202.1 & RBS 0016   \\
3FGL J0010.5$-$1425 & RFC J0010$-$1420 & 00 10 41.940201 & -14 20 20.33807 &   0.26 &  0.56 & -0.149 & 5.49 & 1.24 &  21.0 &    8.4 &            \\
      \hline
    \end{tabular}
        \par\medskip\small
    Column description: 3FGL --- 3FGL identifying name; VLBI Name --- IAU name; Right Ascension/Declination --- J2000 coordinates of VLBI detection; $ \sigma_\alpha $ --- uncertainty in right ascension without $\cos \delta$
    factor in mas; $ \sigma_\delta $  --- uncertainty in declination in mas; Corr --- correlation between right ascension and declination estimates; D --- separation between radio and $\gamma$-ray source in arcminutes; N --- normalized separation between radio and $\gamma$-ray source; S --- total VLBI flux density at 8~GHz integrated over the images in mJy; $\Lambda$  --- likelihood ratio; 3FGL ass -- 3FGL association.
    \label{t:ass}  
\end{table*}

\subsection{Previously reported 3FGL associations\label{sec:known_assoc}}

   Among 170 sources with associations reported by the 3FGL team we overturned
5 associations for radio sources that are AGN and provide alternative 
associations for two pulsars (see Table~\ref{t:known_assoc}). For all 5 AGN 
we suggest new associations that are closer to the 3FGL position than the 
associations suggested by the {\it Fermi} team. \citet{2015arXiv150203251L} 
report $\gamma$-ray pulsation with the same period as millisecond pulsar 
PSR~J0931-19. The pulsar PSR~J0931-19 and the AGN RFC\,J0930-1903 are two different 
objects. One of the explanations why two methods, detection of parsec scale emission 
and detection of radio pulses, led to different sources is that errors of 3FGL\,J0930.9-1904 
\textit{Fermi} localization may be larger than their formal uncertainties.

\begin{table}[h]
    \caption{Objects with alternative associations based on parsec-scale 
             emission.}
    \par\medskip\par
    \centering
    \scriptsize
    \sc
    \hspace{-4em}
    \begin{tabular}{l @{$\:\:\:$} l @{$\:\:$} l @{$\:\:$} l}
       \hline
       3FGL name           & type & old assoc             & suggested assoc.   \\
       \hline
       3FGL J2021.9$+$0630 & agn & 87GB 201926.8$+$061922 &  RFC J2021$+$0629  \\
       3FGL J0332.0$+$6308 & agn & GB6 J0331$+$6307       &  RFC J0331$+$6308  \\
       3FGL J0150.5$-$5447 & agn & PMN J0150$-$5450       &  RFC J0150$-$5450  \\
       3FGL J0415.7$-$4351 & agn & SUMSS J041605$-$435516 &  RFC J0416$-$4350  \\
       3FGL J1838.5$-$6006 & agn & SUMSS J183806$-$600033 &  RFC J1838$-$6005  \\
       3FGL J0930.9$-$1904 & psr & PSR J0931$-$19         &  RFC J0930$-$1903  \\
       3FGL J2017.6$-$1616 & psr & PSR J2017$-$1618       &  RFC J2017$-$1618  \\
    \end{tabular}
    \label{t:known_assoc}  
\end{table}

  We have detected radio counterparts within $1'$ of all 122 observed X-ray
sources associated with {\it Fermi} sources. The median position differences 
between ATCA and X-ray positions is $6.8''$ with 7 objects detected at distances
in a range of $30$--$51''$. ATCA observations provided improvement in positions for 85\% of the sources.
We observed 84 out of 122 sources with the VLBA and detected 82.
Among the remaining 38 objects, 28 have a flux density from ATCA 
observations above 10 mJy. They will be observed with VLBI in the future.

  Based on these observations, we can conclude that association of 3FGL sources 
based on the X-ray emission matches very well with association based on radio
emission from parsec scales. We strongly confirmed two out of three X-ray associations, and
found no X-ray sources from our sample of 122 objects that have not been 
detected at 5.5 or 9~GHz.

\subsection{Extended Sources}

All target fields observed with ATCA and VLA were imaged and visually inspected. 
We found 158 fields where sources showed emission larger than the synthesized beam. 
Two classes of such extended sources were identified: Large Angular Structures (LAS)
that were either resolved (or over-resolved) and double sources or point sources with
extensions. Table~\ref{tab:extended} lists the 3FGL source in those fields and their 
corresponding localization within that field, distance to the 3FGL position,
and structural classification.

\begin{figure}
 \includegraphics[width=0.48\textwidth]{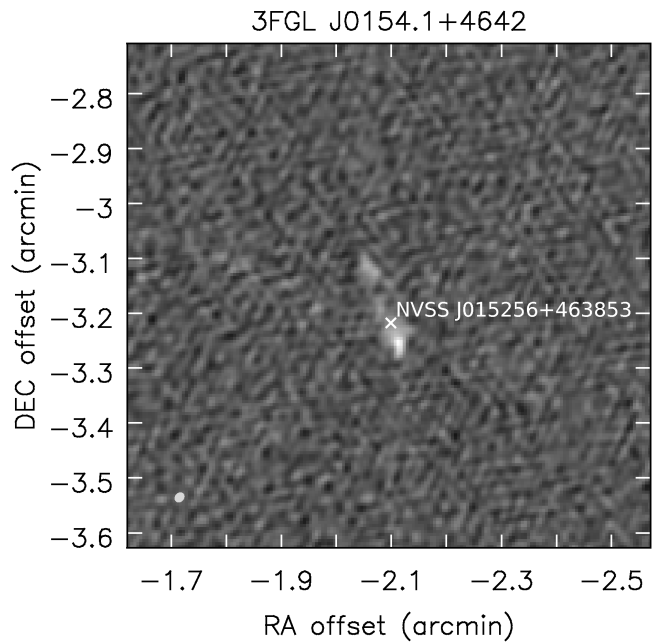}
 \caption{This extended source was found in the field of 3FGL J0154.1+4642 observed with the VLA
          on 2015 Mar. 16. The 
          coordinates are shown relative to the centroid of the $\gamma$-ray 
          localization, with the field shown lying inside the 68\% and 95\% localization
          confidence areas. The peak flux in the 5.0 GHz map is 0.6\,mJy\,beam$^{-1}$ , the integrated
          flux density is 4.6 mJy, and the image rms at the center of the field of view is 58\,$\mu$Jy\,beam$^{-1}$. The in band spectral index at the peak value
          is $-$2.31$\pm$0.69. The position of NVSS J015356+463853 is marked with an 
          'x' in the field. The filled circle in the 
            lower left corner indicates the size of the synthesized beam.\label{fig:J0154+4642}}
\end{figure}

\begin{figure}
 \includegraphics[width=0.48\textwidth]{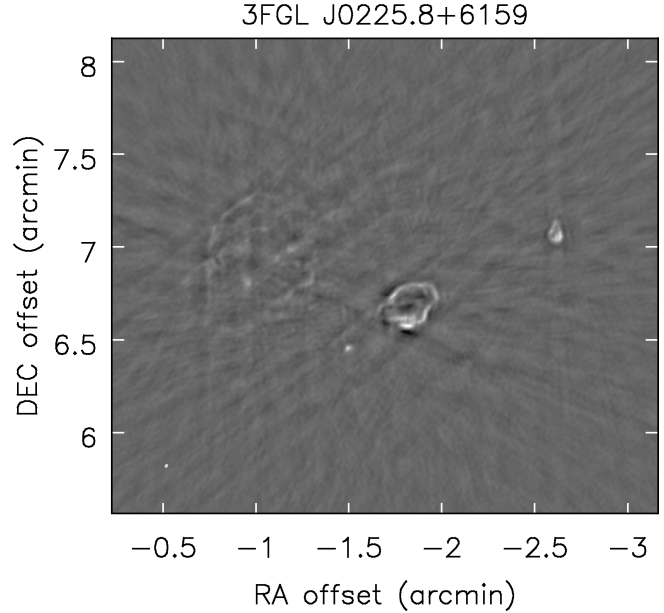}
 \caption{These extended sources were found in the field of 3FGL J0225.8+6159 observed with the VLA on
          2015 Mar. 16. 
          The coordinates are shown relative to the centroid of the $\gamma$-ray 
          localization, with the field shown lying far outside the 95\% localization
          confidence region. The peak flux of the central source shown in the 5.0 GHz 
          map is 6.3\,mJy\,beam$^{-1}$, the integrated flux density is 50.0 mJy, and the 
          image rms at the center of the field of view is 0.1\,mJy\,beam$^{-1}$. The peak flux density of the North
          West source is 5.1\,mJy\,beam$^{-1}$ with a flux density of 30.7 mJy.
          For both the in band spectral index is $-$4$\pm$1. The filled circle in the 
            lower left corner indicates the size of the synthesized beam.\label{fig:J0225+6159}}
\end{figure}

\begin{figure}[htbp!]
 \includegraphics[width=0.48\textwidth]{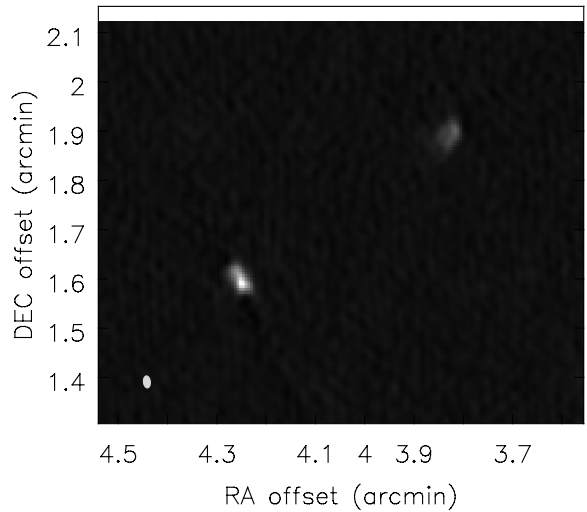}
 \caption{This extended source was found in the field of 3FGL J1839.9+7646 with the VLA
          at 5.0\,GHz observed on 2015 Mar. 16. The coordinates are shown relative to the center of the $\gamma$-ray 
          localization, with the field shown lying inside the 68\% and 95\% localization
          confidence areas. The in band spectral index of the South-East source is $-$2.5$\pm$0.1 and that of 
          the North-West source is $-$1.9$\pm$0.2. The peak flux in the map is 
          6.2\,mJy\,beam$^{-1}$, the integrated flux density over both sources
          is 26.6\,mJy, and the image rms at the center of the field of view is 65\,$\mu$Jy\,beam$^{-1}$. The filled circle in the 
            lower left corner indicates the size of the synthesized beam.\label{fig:J1839+7646}}
\end{figure}

\begin{figure}
 \includegraphics[width=0.48\textwidth]{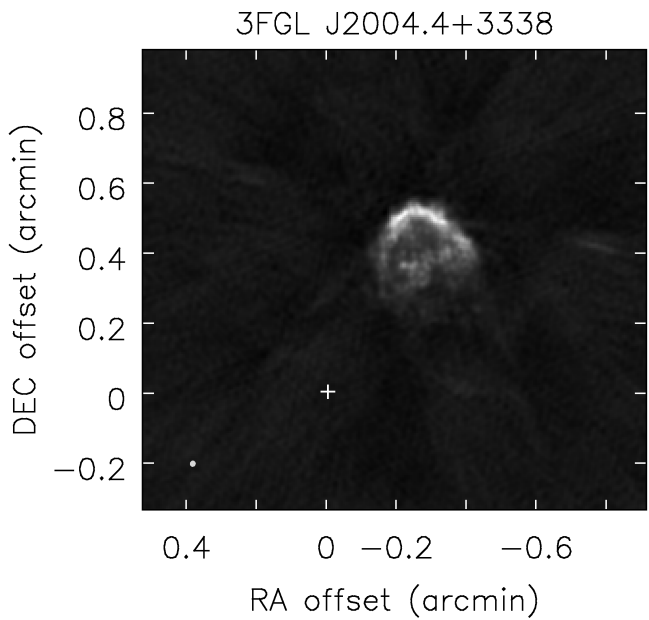}
 \caption{This shell-like structure was found with the VLA at 5.0\,GHz in the field near 3FGL\,J2004.4+3338
          observed on 2015 Mar 16.
          The coordinates are shown relative to the centroid of the $\gamma$-ray 
          localization, which is marked with a cross. 
          The object lies within the 68\% and 95\% confidence region of the $\gamma$-ray
          source. It has an in band spectral index of $-$0.9$\pm$0.3, an integrated
          flux density of 247\,mJy, a peak flux of 4.2\,mJy\,beam$^{-1}$, and an image rms at the center of the 
          field of view of 70\,$\mu$Jy\,beam$^{-1}$. The filled circle in the 
            lower left corner indicates the size of the synthesized beam.\label{fig:J2004+3338}}
\end{figure}

\begin{figure}
 \includegraphics[width=0.48\textwidth]{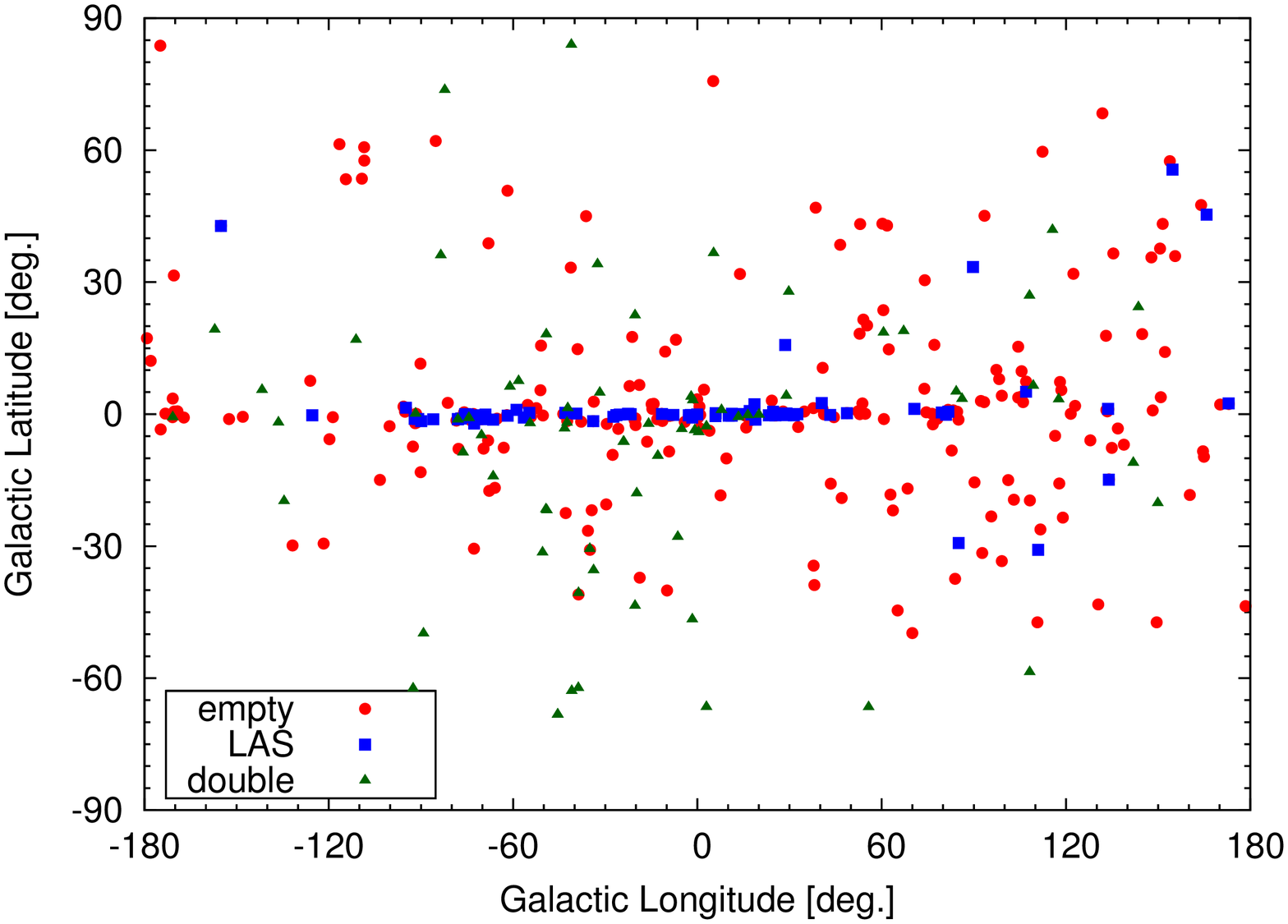}
 \caption{Sky distribution in Galactic coordinates of empty fields, large angular 
          size sources, and fields with double or extended point sources.\label{fig:distribution}}
\end{figure}

Most sources found are either sources exhibiting a shell like structure, as shown
in Figures~\ref{f:aofus_ima1}, \ref{fig:J0225+6159}, and 
\ref{fig:J2004+3338}, while a second group of 
sources shows a morphology that resembles double lobed sources or one-sided
jets, as in Figures~\ref{fig:J0154+4642} and \ref{fig:J1839+7646}. In 36 cases 
an extended source was found within the 2$\sigma$ \textit{Fermi} localization error ellipse, 
of which 19 were classified as LAS and 17 as double source. Fig.~\ref{fig:distribution}
shows the sky distribution of the extended sources. From this plot it is evident
that LAS sources are primarily located within the Galactic plane, whereas double
sources are primarily found outside of the Galactic plane. Judging from the morphologies 
and radio spectral properties the objects shown in
Figures~\ref{fig:J0225+6159} and \ref{fig:J2004+3338} we see that they resemble supernova remnant 
shells or HII regions. 
\begin{itemize}
 \item In the case of 3FGL J2004.4+3338 the shell is found within 
the 1-$\sigma$ localization of the $\gamma$-ray source and is most likely associated 
with such. An infrared counterpart is found for this structure in the ALLWISE catalog \citep{2010AJ....140.1868W,2011ApJ...731...53M},
J200423.63+333904.2.
 \item In the case of 3FGL\,J0225.8+6159 the shells are found more than 5-$\sigma$ away
from the $\gamma$-ray localization. This object also has an infrared (IR) counterpart in the
ALLWISE catalog, J022537.37+620553.3. This indicates the presence of a HII region
in both cases.
 \item The case of Figure~\ref{fig:J0154+4642} shows an extended steep spectrum radio 
source that could be a compact symmetric object \citep{1994ApJ...432L..87W,1996ApJ...460..634R}, but more likely is a radio galaxy, due
to the lack of evidence for the presence of a single compact core. 
Higher resolution observations are required to understand the nature of this object.
It is found within the 2-$\sigma$ localization of 3FGL\,J0154.1+4642.
 \item Figure~\ref{fig:J1839+7646} shows an example of two extended sources separated by 
$\sim$8$''$. Both of these have a steep spectral index and if related resemble the 
structure of a radio galaxy. In this case this structure is located within 
1-$\sigma$ of the $\gamma$-ray localization.  
\end{itemize}

Overall, the above defined source classes seem to separate a Galactic and an 
extragalactic population of extended radio sources.

\begin{table}
        \centering
	\caption{First eight rows of 158 extended sources found in radio observations
                 targeting 3FGL unassociated $\gamma$-ray objects. The full table is available 
                 in the electronic attachment.\label{tab:extended}}
	\par\medskip\par
	\begin{tabular}{llllll}
                \hline
                \small
		3FGL Name & RA  & Dec & D      & Nd & C\\
                          & hms & dms & $'$ &   \\
                \hline
                J0000.1$+$6545  & 23 59 41.68 & $+$65 42 3.456 & 4.14 & 3.3 & 2\\
                J0003.2$-$5246  & 00 03 11.60 & $-$52 41 37.20 & 5.04 & 5.5 & 2\\
                J0003.4$+$3100  & 00 03 20.13 & $+$30 55 25.01 & 5.24 & 2.1 & 1\\
                J0020.9$+$0323  & 00 20 50.30 & $+$03 23 43.10 & 1.29 & 1.6 & 2\\
                J0026.2$-$4812  & 00 25 37.47 & $-$48 16 39.11 & 7.27 & 4.5 & 2\\
                J0154.1$+$4642  & 01 53 56.09 & $+$46 38 51.47 & 3.89 & 1.7 & 1\\
                J0225.8$+$6159  & 02 25 37.82 & $+$62 05 53.76 & 6.92 & 5.8 & 1\\
                J0026.2$-$4812  & 00 25 53.27 & $-$48 16 43.19 & 5.34 & 3.3 & 2\\
                \hline
	\end{tabular}
   	\begin{flushleft}
    	 Column description:\\
     		3FGL -- Object identifier\\
     		RA   -- Right Ascension\\
     		Dec  -- Declination\\
                D -- Distance from $\gamma$-ray localization\\
                Nd --  $\sigma$ from $\gamma$-ray localization\\
                C -- Extended source class, 1 -- large angular 
                         scale object; 2 -- double or jetted source
   \end{flushleft}        
\end{table}

\subsection{Empty Fields}

The list of 245 3FGL sources with no radio counterpart brighter than 2~mJy at 5\,GHz within 
the 3-$\sigma$ \textit{Fermi} localization error ellipse is presented in Table~\ref{t:empty}. Of these, 237 are 
marked as unassociated in the 3FGL catalog, four are associated with pulsars, one is associated 
with NVSS J224604+154437. The latter association was not confirmed. The remaining three are 
associated with supernova remnants, namely 3C\,391, MSH 17-39 and 3CTB 37A. Among those 245 ``empty'' 
fields we found evidence for extended radio emission in 39 cases listed in Table~\ref{tab:extended}. 
In those cases there is evidence for extended emission with a total flux density of $>$2\,mJy, however no individual
bright spots exceed $>$2\,mJy. This leaves 206 empty fields in which we did not detect a radio source above 2~mJy 
with the VLA and ATCA. We call these fields ``empty'' in the sense there are no radio sources brighter than 2~mJy
within the observed frequency range. As \citet{2016arXiv160603450F} noted, 
some of these fields contain weak objects listed in low-frequency NVSS and SUMSS catalogs.
The locations of the empty fields on the sky are illustrated in Fig.~\ref{fig:distribution},
which shows no apparent concentration of empty fields in the Galactic plane. 
However, counting the number of empty fields, 27\% (65/245) lie within 1$^\circ$ of 
the Galactic plane. There were 54 matches of TGSS ADR1 sources \citep{2016arXiv160304368I} that were found
within 3$\sigma$ of the $\gamma$-ray localization. Such sources are most likely
steep spectrum radio sources and are good candidates for Pulsar searches. 

Notable are 95/245 (39\%) empty fields that are found $\pm$10$^\circ$ outside of the 
galactic plane. Their median $\gamma$-ray flux density is 
$5.96\times10^{-13}$ph\,cm$^{-2}$\,s$^{-1}$\,MeV$^{-1}$ and median $\gamma$-ray 
spectral index is 2.4. Their median variability index is 47.0 \citep{2011ApJ...743..171A}, which indicates 
that most of the sources can be assumed to be non-variable. The spectral index 
corresponds to the median of the total population of $\gamma$-ray point sources.

\begin{table}
    \caption{The first 8 rows of the list of 245 3FGL unassociated 
             objects that have no radio source brighter 2~mJy within 
             the 3-$\sigma$ \textit{Fermi} localization error ellipse. In the 98 cases where 
             a TGSS ADR1 was found within the 3-$\sigma$ \textit{Fermi} localization
             error ellipse, the corresponding identifier, separation, 
             and $\sigma$ are listed.}
    \par\medskip\par
    \centering
    \begin{tabular}{llll}
	   \hline
	   3FGL & TGSS ADR1 & D & Nd\\
                &  & ($'$) & \\
           \hline
3FGL J0001.6+3535 &  J000051.8+352741 & 12.0 & 150.6 \\
3FGL J0003.5+5721 &  J000332.3+572711 & 05.6 & 167.9 \\
3FGL J0004.2+6757 &  -                & -    &       \\  
3FGL J0017.1+1445 &  -                & -    &       \\
3FGL J0022.7+4651 &  J002212.5+464402 & 09.0 & 122.5 \\
3FGL J0032.5+3912 &  -                & -    &       \\
3FGL J0039.3+6256 &  -                & -    &       \\
3FGL J0051.6+6445 &  -                & -    &       \\
           \hline
    \end{tabular}
    \label{t:empty}
    
    Column descriptions: 3FGL -- 3FGL catalog source name; TGSS ADR1 -- TGSS 
                          ADR1 source name \citep{2016arXiv160304368I}; D -- 
                          angular distance in arcmin between $\gamma$-ray and 
                          TGSS ADR1 localization; Nd -- normalized angular 
                          distance between $\gamma$-ray and TGSS ADR1 localization.
\end{table}
\par\vspace{3ex}\phantom{A}\par 

\section{Discussion \& Summary} \label{sec:steady} \label{sec:discussion}

With the newer \textit{Fermi}/LAT catalogs providing longer averaging time one can see that
from 2FGL to 3FGL the median fluxes of unassociated $\gamma$-ray sources have 
dropped by a factor of 1.8. Since the two catalogs 2FGL and 3FGL were derived using different
diffuse background models and slightly different methods, we use in the following 
analysis only values from 3FGL and when referring to 2FGL sources the values of the 300
cross-listed objects are chosen. Comparing the group of previously found unassociated
$\gamma$-ray sources with the group of newly found ones and using the 3FGL reported
variability indices we find that 3.0\% of previously reported objects are variable
compared to 2.4\% of newly found objects. This shows that the majority of 
unassociated $\gamma$-ray sources do not show significant $\gamma$-ray variability 
and could be considered steady $\gamma$-ray emitters. Since most of the unassociated
sources are steady emitters this cannot be used to distinguish different populations
such as fields associated with AGN or empty fields. An additional complication arises
from the fact that finding variability in fainter $\gamma$-ray sources is very 
difficult given the poor photon statistics. 

Despite pushing toward lower $\gamma$-ray flux densities in the newest $\gamma$-ray
source catalogs we continue to increase the number of AGN associations using radio
observations. Eventually, this will lead to a reduction of the radio flux density completeness limit of 
$\gamma$-ray radio counterparts by a factor of 10, from 100\,mJy to around 10\,mJy. 
With the results presented here, together with previous observations, we have firmly 
associated 144 new $\gamma$-ray objects with newly discovered AGN and improved 
positions of 170 previously associated sources to the milliarcsecond level of accuracy.
We expect to increase this number after all VLBI follow-up observations have been
completed. This marks the single largest reduction in unassociated sources by any 
follow-up program to date and demonstrates the strength of this effort in 
comparison to solely relying on the analysis of existing multi-wavelength 
catalogs. In the global context this corresponds to a reduction of 3FGL unassociated 
sources by 14\%, leaving 901 sources without a multi-wavelength counterpart. Note that
the total number of known $\gamma$-ray pulsars is currently 
220\footnote{\url{https://confluence.slac.stanford.edu/display/GLAMCOG/Public+List+of+LAT-Detected+Gamma-Ray+Pulsars}},
against 1550 $\gamma$-ray AGN with radio emission from parsec scales. A recent 
publication discussing potential new $\gamma$-ray pulsar candidates 
highlights the challenges involved in finding more $\gamma$-ray pulsars 
\citep{2016arXiv160603450F}.   

The most remarkable result presented here are the so-called empty fields, which are
not necessarily empty, but for which no radio point source brighter than
several mJy was found in the region of the $\gamma$-ray localization. We first 
pointed these fields out in \citet{2015ApJS..217....4S}. For 2FGL unassociated 
sources those empty fields were primarily localized within the galactic plane. 
However, 39\% of empty fields were found outside of the galactic plane. This hints 
at an extragalactic distribution in addition to the previously found galactic one. 
There is still a possibility that 
a large number of empty fields are related to pulsars, which are also found 
outside of the galactic plane. However, another possibility could be the presence 
of radio quiet AGN that are not found in optical/IR catalogs. It is expected that
high-frequency peaked AGN have very faint radio flux densities, due to the 
shifting of their overall spectral energy distribution to higher energies.

We found in at least 16\% (154) of observed radio fields extended radio sources, of which
36 lie within 2$\sigma$ of the $\gamma$-ray source. These extended sources were 
broken down into two sub-classes, one indicating a large angular structure object
and the second indicating the presence of a double or jet-like extended source. 
The first subclass could be related to galactic SNR or HII regions, which should
be further investigated with more compact interferometer configurations or radio
observations at lower frequencies, as well as a systematic comparison to infrared
observations which could distinguish the SNR origin from HII regions. The majority
of sources found in the second subclass are most likely radio galaxies, but could
also contain compact symmetric objects (CSOs). High resolution VLBI spectral imaging 
observations are needed to determine the nature of those objects, which has been
done in the case of {\sf 2234$+$282} \citep{2016AN....337...65A}.  However no convincing
CSO $\gamma$-ray emitter has been found to date at redshifts beyond z$>$0.006.
Establishing a firm association of a CSO or distant radio galaxy with a $\gamma$-ray source would
challenge existing interpretations of the physics of the production of $\gamma$-ray 
emission. 

In summary, we present an update of our work on new associations for $\gamma$-ray
unassociated sources found by \textit{Fermi}. For all 3FGL unassociated sources we
performed radio observations with compact interferometers, which provided 2097 radio
objects as candidates for association. We have observed a large fraction of those providing
milliarcsecond scale detections. For 144 of those detections we provide firm 
associations of newly found AGN based on statistical likelihood. More remarkably, 
we found a new population of $\gamma$-ray fields devoid of any compact radio source 
brighter than 2\,mJy that lies outside of the galactic plane. We also provide a 
list of fields for which we found large scale radio structures that are indicative 
of SNRs, HII regions, and radio galaxy counterparts. 

High resolution radio follow-up observations are underway to cover all newly
reported association candidates with which we expect to double the current number of new
AGN associations. 

\acknowledgments

We thank the anonymous referee for a thorough review which improved the quality of this publication.
FKS, GBT, and LP acknowledge support by the NASA Fermi Guest Investigator 
program, grants NNX12A075G, NNX14AQ87G and NNX15AU85G. The National Radio 
Astronomy Observatory is a facility of the National Science Foundation operated under cooperative
agreement by Associated Universities, Inc. The Australia Telescope
Compact Array \& Long Baseline Array are part of the Australia
Telescope National Facility which is funded by the Commonwealth of
Australia for operation as a National Facility managed by CSIRO. The authors 
also thank Jamie Stevens and Elaine Sadler with their support of the ATCA observations.
FKS dedicates this paper to his wife M\'onica for her support and understanding
leading to this publication.

This publication makes use of data products from the Wide-field Infrared Survey 
Explorer, which is a joint project of the University of California, Los Angeles, 
and the Jet Propulsion Laboratory/California Institute of Technology, and NEOWISE, 
which is a project of the Jet Propulsion Laboratory/California Institute of Technology. 
WISE and NEOWISE are funded by the National Aeronautics and Space Administration.
This work made use of the Swinburne University of Technology software
correlator, developed as part of the Australian Major National
Research Facilities Programme and operated under licence. This
research has made use of NASA's Astrophysics Data System and has made
use of the NASA/IPAC Extragalactic Database (NED) which is operated by
the Jet Propulsion Laboratory, California Institute of Technology,
under contract with the National Aeronautics and Space
Administration. This research has made use of data, software and/or
web tools obtained from NASA's High Energy Astrophysics Science
Archive Research Center (HEASARC), a service of Goddard Space Flight
Center and the Smithsonian Astrophysical Observatory, of the SIMBAD
database, operated at CDS, Strasbourg, France, and the TOPCAT software
version
4.1\footnote{\url{http://www.star.bris.ac.uk/\~mbt/topcat/}}
\citep{2005ASPC..347...29T}. The authors made use of the database CATS 
\citep{2007HiA....14..636V} of the Special Astrophysical Observatory.



\vspace{5mm}
\facilities{VLA,ATCA,VLBA,LBA,Fermi/LAT}




\bibliography{references}



\end{document}